\documentclass{new_tlp}

\usepackage{amsmath}

\newcommand{\limp}{\rightarrow}
\newcommand{\init}{\mathit{init}}
\newcommand{\next}{\mathit{next}}
\newcommand{\safe}{\mathit{safe}}
\newcommand{\inv}{\mathit{inv}}
\newcommand{\binv}{\mathit{binv}}
\newcommand{\lfalse}{\mathit{false}}
\newcommand{\round}{\mathit{round}}
\newcommand{\wf}{\mathit{wf}}
\newcommand{\final}{\mathit{final}}
\newcommand{\io}{\mathit{io}}
\newcommand{\sys}{\mathit{sys}}
\newcommand{\env}{\mathit{env}}
\newcommand{\goal}{\mathit{goal}}

\title{(Quantified) Horn Constraint Solving for Program Verification and Synthesis}

\author{Andrey Rybalchenko\\
  Microsoft Research\\
  \email{rybal@microsoft.com}
}

\begin{document}

\maketitle

\begin{abstract}
We show how automatic tools for the verification of linear and branching time properties of procedural, multi-threaded, and functional programs as well as program synthesis can be naturally and uniformly seen as solvers of constraints in form of (quantified) Horn clauses over background logical theories. Such a perspective can offer various advantages, e. g., a logical separation of concerns between constraint generation (also known as generation of proof obligations) and constraint solving (also known as proof discovery), reuse of solvers across different verifications tasks, and liberation of proof designers from low level algorithmic concerns and vice versa.

\noindent
To appear in Theory and Practice of Logic Programming (TPLP)
\end{abstract}


\section{Introduction}

A variety of interesting and important verification and synthesis
questions about programs can be formulated as constraint satisfaction
problems in form of implication/Horn constraints over a suitable
background
theory~\cite{ThreaderPOPL2011,ThreaderCAV2011,GrebenshchikovPLDI12,ThreaderTerm12,SMT2012,CinderellaPOPL14}.
Verification of temporal properties and program synthesis are
particularly relevant examples of such questions.
Then, given an efficient constraint solver we can obtain a program
verifier or synthesizer by composing it with a constraint generator. 
As a result we achieve a separation of concerns such that deduction
rules for reasoning about programs can be developed independently and
interact compositionally with inference engines (that automate the
deduction process).

\section{Constraint generation}

We illustrate the constraint based approach using the following example.
Let a program be represented by an assertion $\init(v)$ that describes
a set of initial states, and an assertion $\next(v, v')$ that
describes a binary transition relation.

\paragraph{Safety}

To prove that each state that is reachable from an initial state by
following the transition relation satisfies an assertion $\safe(v)$ we
generate the following constraint.
\begin{equation*}
  \begin{array}[t]{@{}l@{}}
    \exists \inv: \\[\jot]
    \qquad
    \begin{array}[t]{@{}l@{\;}l@{}}
      & (\forall v: \init(v) \limp \inv(v)) \\[\jot]
      \land & 
      (\forall v\; \forall v': \inv(v) \land \next(v, v') \limp \inv(v')) \\[\jot]
      \land & 
      (\forall v: \inv(v) \limp \safe(v))
    \end{array}
  \end{array}
\end{equation*}
Here we rely on second order existential quantifier to model the
search for a safe (forward-)inductive  invariant.
Any model of $\inv$ is such an invariant.

Note that by using the following constraint system (but not the
solver/inference engine) we can change the proof rule from forward
invariance to backward invariance.
\begin{equation*}
  \begin{array}[t]{@{}l@{}}
    \exists \binv: \\[\jot]
    \qquad
    \begin{array}[t]{@{}l@{\;}l@{}}
      & (\forall v: \neg\safe(v) \limp \binv(v)) \\[\jot]
      \land & 
      (\forall v\; \forall v': \binv(v') \land \next(v, v') \limp \binv(v)) \\[\jot]
      \land & 
      (\forall v: \binv(v) \land \safe(v) \limp \lfalse)
    \end{array}
  \end{array}
\end{equation*}
Furthermore, we can combine forward and backward reasoning in the same
constraint system:
\begin{equation*}
  \begin{array}[t]{@{}l@{}}
    \exists \inv\;\exists \binv: \\[\jot]
    \qquad
    \begin{array}[t]{@{}l@{\;}l@{}}
      & (\forall v: \init(v) \limp \inv(v)) \\[\jot]
      \land & 
      (\forall v\; \forall v': \inv(v) \land \next(v, v') \limp \inv(v')) \\[\jot]
      \land & (\forall v: \neg\safe(v) \limp \binv(v)) \\[\jot]
      \land & 
      (\forall v\; \forall v': \binv(v') \land \next(v, v') \limp \binv(v)) \\[\jot]
      \land & 
      (\forall v:  \inv(v) \land \binv(v)  \limp \lfalse)
    \end{array}
  \end{array}
\end{equation*}
All these reasoning approaches can be automated by the same solver,
which can be made highly beneficial through consolidation of
heuristics, optimizations, and improvements.

\paragraph{Termination}

To prove program termination, i.e., absence of infinite sequences of
states that start in an initial state and follow the transition
relation, we can resort to the following constraint.
\begin{equation*}
  \begin{array}[t]{@{}l@{}}
    \exists \inv\; \exists \round: \\[\jot]
    \qquad
    \begin{array}[t]{@{}l@{\;}l@{}}
      & (\forall v: \init(v) \limp \inv(v)) \\[\jot]
      \land & 
      (\forall v\; \forall v': \inv(v) \land \next(v, v') \limp \inv(v')) \\[\jot]
      \land & (\forall v: \neg\safe(v) \limp \binv(v)) \\[\jot]
      \land & (\forall v\;\forall v' : \inv(v) \land \next(v, v') \limp \round(v, v')) \\[\jot]
      \land & \wf(\round)
    \end{array}
  \end{array}
\end{equation*}
Note that we rely on a second order predicate $\wf$ that holds for
well-founded relations. 
Each solver can rely on a specific way of proving well-foundedness,
e.g., using abstract interpretation~\cite{CousotCousot-POPL12},
ranking functions~\cite{Turing49} or transition
invariants~\cite{LICS04,Thesis}.

Instead of using forward invariance to keep track of reachable states,
we can use alternative approaches, as exemplified by the constraint
systems for reasoning about program safety.

\paragraph{Information flow property}

Security properties often require proving that any alternation of
a (secret) program input is undetected by observing its (public)
output.
Let $\final(v)$ be an assertion that describes the set of all states
where computation can stop.
The program satisfies the non-interference property if the following
constraint is satisfiable.
\begin{equation*}
  \begin{array}[t]{@{}l@{}}
    \exists \io: \\[\jot]
    \qquad
    \begin{array}[t]{@{}l@{\;}l@{}}
      & (\forall v: \init(v) \next(v, v')\limp \io(v, v')) \\[\jot]
      \land & 
      (\forall v\; \forall v'\;\forall v'': \io(v, v') \land \next(v', v'') \limp \io(v, v'')) \\[\jot]
      \land & (\forall v\;\forall v'\;\forall w\;\forall w' : v\neq w \land \io(v, v') \land \io(w, w') \land \final(v) \land \final(w') \limp w=w') 
    \end{array}
  \end{array}
\end{equation*}
This constraint captures reasoning about pairs of computations through
self-composition of input/output relation of the program, instead of
performing a source-to-source transformation~\cite{SelfComp} or a
specialized inference procedure~\cite{Leak2009}.

\paragraph{Temporal property}

For proving properties expressed in temporal logics we can rely on
existing proof systems that reduce temporal reasoning to first-order
reasoning (with auxiliary assertions), e.g., the proof system for
CTL*~\cite{KestenTCS95}.
For example, to prove that there exists a computation that visits
states satisfying an assertion $p(v)$ until finally it reaches a state
satisfying an assertion $q(v)$, i.e., $(\init(v), \next(v, v'))
\models E\;(p(v) U q(v))$, we need to solve the following constraint.
\begin{equation*}
  \begin{array}[t]{@{}l@{}}
    \exists \inv\; \exists \round: \\[\jot]
    \qquad
    \begin{array}[t]{@{}l@{\;}l@{}}
      & (\forall v: \init(v) \limp \inv(v)) \\[\jot]
      \land & 
      (\forall v: \inv(v) \land \neg q(v)  \limp \exists v': \next(v,v') \land \inv(v') \land \round(v, v')) \\[\jot]
      \land & (\forall v: \inv(v) \limp p(v)) \\[\jot]
      \land & \wf(\round)
    \end{array}
  \end{array}
\end{equation*}
Note that we model the existence of a computation by relying on
existential quantification in the constraints together with a
recursive dependency~\cite{ehsf}. 

\paragraph{Reactive synthesis}

We formulate the reactive synthesis problem as constraint solving by
turning this problem into a game solving
problem~\cite{CinderellaPOPL14}.
For example, consider the synthesis of a system with an unknown
transition relation $\sys(v, v')$ that is executed in an adversarial
environment with a given transition relation~$\env(v, v')$. 
The system's objective is to reach a state satisfying an assertion
$\goal(v)$ regardless of environment's behavior.
The following constraint characterizes the synthesis problem.
\begin{equation*}
  \begin{array}[t]{@{}l@{}}
    \exists \inv\; \exists \round: \\[\jot]
    \qquad
    \begin{array}[t]{@{}l@{\;}l@{}}
      & (\forall v: \init(v) \limp \inv(v)) \\[\jot]
      \land & 
      (\forall v\;\forall v': \inv(v) \land \neg \goal(v) \land \env(v, v') \limp 
      \exists v'': 
      \begin{array}[t]{@{}l@{}}
        \sys(v',v'') \land \inv(v'')\\[\jot]
        \mathrel{\land} \round(v, v'')) \\[\jot]
      \end{array}\\
      \land & \wf(\round)
    \end{array}
  \end{array}
\end{equation*}
Other temporal objectives can be satisfied by relying on a temporal
proof system, akin to temporal verification.

\section{Constraint solving}

In the previous section we illustrated how program verification and
synthesis question can be formalized as second-order constraint
solving problems in form of recursive implication constraints with
well-foundedness conditions, and quantifier alternation. 

Solving (various sub-classes of) such constraints is a thriving area
of research.
Solvers often take advantage of the fact that constraints have
Horn-like structure, which enables iterative, abstraction based
solving approaches.

Important classes include recursion-only
case~\cite{muz,GrebenshchikovPLDI12,Duality,RummerDisj2013} that is
facilitates reasoning about safety properties, extension with
well-foundedness for reasoning about liveness
properties~\cite{GrebenshchikovPLDI12}, extension with universal
quantification for inferring universally quantified
invariants~\cite{UnivSAS2013}, and extension with existential
quantification for dealing with synthesis and branching time
questions~\cite{ehsf}.
Often such solvers rely on recursion-free fragments of implication
constrains~\cite{McMillanSplitProver,HornLIUIF11,InterSystems13,TreeVampire13}
and ranking function synthesis~\cite{RybalPodelskiVMCAI04} as a basic
inference components.

Constraint logic programming offers an effective tool for implementing
solvers for quantified (Horn) implication constraints with
well-foundedness, e.g., the HSF solver and its
extensions~\cite{HSF,ehsf,UnivSAS2013} is implemented using ideas of
blending meta-logic programming and constraint logic
programming~\cite{RybalPodelskiPADL07}.

\bibliographystyle{acmtrans} 
\bibliography{biblio}

\end{document}